\documentclass[%
 preprint,
 amsmath,amssymb,
 aps,
]{revtex4-2}
\usepackage{caption}
\usepackage{subcaption}
\usepackage{rotating}
\usepackage{graphicx}
\usepackage{dcolumn}
\usepackage{bm}
\usepackage{braket}
\usepackage{fancybox}
\usepackage{tikz}
\usetikzlibrary{trees,arrows}

\begin{document}

\preprint{APS/123-QED}

\title{Wave function realization of a thermal collision model}
\thanks{A footnote to the article title}%

\author{Uriel Shafir}%
 \email{uriel.shafir@mail.huji.ac.il}
\affiliation{The Institute of Chemistry, The Hebrew University of Jerusalem, Jerusalem 9190401, Israel}%
\author{Ronnie Kosloff}%
\email{kosloff1948@gmail.com}
\affiliation{The Institute of Chemistry, The Hebrew University of Jerusalem, Jerusalem 9190401, Israel}

\date{\today}

\begin{abstract}
An efficient algorithm to simulate dynamics of open quantum system is presented. The method describes the dynamics by unraveling stochastic wave functions converging to a density operator description. The stochastic techniques are based on the quantum collision model.
Modeling systems dynamics by wave functions and modeling the interaction with the environment with a collision sequence reduces the complexity scale significantly. The algorithm developed, can be implemented on quantum computers.
We introduce stochastic methods that exploit statistical characters of the model, as Markovianity, Brownian motion and binary distribution. The central limit theorem is employed to study the convergence of distributions of stochastic dynamics of pure quantum states represented by wave vectors. By averaging a sample of functions in the distribution we prove and demonstrate the convergence of the dynamics to the mixed quantum state described by a density operator.

\end{abstract}

\maketitle


\newpage

\section{\label{sec:introduction}introduction}

In reality every quantum system  is open, while an isolated system is an exception. Therefore the main setback in simulating and modeling  real life quantum system is the high cost in computation.
To analyse the cost, we will first describe the generic description of an open quantum system, dive into description of the dynamics and observe the computation problem.
After reviewing the current methods we will demonstrate a scheme able to lower the computation complexity.

An open quantum system is generically described by the following:
the systems' Hamiltonian and systems density matrix(DM) -$\hat H_s$, $\hat \rho_s$
the measuring apparatus - M
and the environment Hamiltonian $\hat H_{B}$.
Contemporary examples are the IBM and Google quantum machines,
the nascent steps toward quantum computing \cite{arute2019quantum}.
If these devices are left alone, the quantum system -$\hat \rho_s$, will reach thermal equilibrium with the extremely cold surrounding temperature $T_{B}$. The device is assembled from quantum circuits, where the computation output is measured by a measurement apparatus M.
Another example is the NV center in diamond \cite{doherty2012theory}.
The systems is constructed from a Nitrogen impurity adjecent to a negative vacancy
in  diamond.
The primary quantum system Hamiltonian-$H_s$ is a spin triplet.
The neighboring environment is composed of spins such as other nitrogen atoms or carbon isotopes with nuclear spin ($^{13}C$) in addition to the fluctuations of the vacuum phonons.
A measurement apparatus is being coupled to the NV center able to measure changes in population; the measurement apparatus- M \cite{doherty2013nitrogen}.

Open system dynamics address a system interacting with the environment from the system's perspective. Different approaches have been employed to construct
reduced descriptions in terms of the system observables.
Starting with Bloch a dynamical derivation
based on the weak system bath coupling has led to the quantum Master equation \cite{wangsness1953dynamical,redfield1957theory}.
An alternative  mathematical formulation employing quantum dynamical semigroups has led to a general structure 
termed the Gorini-Kossakowski-Lindblad-Sudarshan (GKLS) equation \cite{lindblad1976generators,gorini1976completely}. Davis has connected the 
perturbation derivation to the general structure \cite{davies1974markovian,alicki2018introduction}.
Non Markovian formulations including memory effects have also been suggested \cite{breuer2002theory}.

Our mission is to develop a simulation algorithm of an open quantum system.
In this approach the system is viewed from a thermodynamic perspective. 
Recent studies by Kosloff and Dann \cite{dann2021open} have paved the ground to the conditions that create consistency between Markovian dynamics and thermodynamic principles in open quantum systems. The consistency conditions are formulated by  
a set of axioms which can be applied to the present study.

The developed quantum simulation  
should be applicable on 
both classical and quantum processors.
This means that the method should be based on a wave functions formulation.
On classical computers the
wave function description is computationally preferable to a density operator description.

When simulating quantum dynamics,
the computational resources of memory scale at least polynomially 
with the size of the Hilbert space.
In addition  every degree of freedom in the system increases the size of the Hilbert space exponentially.
For example in a system of spin particles the size of the Hilbert space is $2^n$, where n is the number of particles.
Describing the system by density matrix squares the memory resources of the computation. Solving for the dynamics
the scaling of the number of operations is the product of the size of the Hilbert space multiplied by the product of propagation time and energy range.
The energy range is also doubled in a density operator description \cite{berman1991time}.
Another advantage in a wavefunction formulation is a pedagogical one. Describing the dynamics of single pure state gives an intuitive sense of the process.

The starting point to describe the dynamics is a global approach which includes the system and environment.
Typically one constructs a global Hamiltonian $\hat H=\hat H_S+\hat H_B +\hat H_{SB}$. Solving for the combined system is computationally prohibited since it consisted of enormous amounts of degrees of freedom. 
To overcome the almost impossible obstacle the whole setup is divided between system and environment.
It is customary to diagonalize  the bath to orthogonal modes either harmonic or composed of an ensemble of spins \cite{prokof2000theory}.
The next step is to obtain an effective reduced equations of motion for the system where the bath enters implicitly.
Assuming an initial uncorrelated system and bath $\hat \rho=\rho_S \otimes \hat \rho_B$ results in a completely positive trace preserving dynamical map (CPTP) 
describing the system propagation $\hat \rho_S(t) =\Lambda_t \hat \rho_S(0)$ \cite{kraus1974operations}. 
Imposing in addition the condition
of Markovianity, Gorini, Kossakowski, Sudarshan and Lindblad 
obtained the general form of the Master equation \cite{lindblad1976generators,gorini1976completely}. The GKLS master equation is describing the dynamics of the open system under the assumption that the bath is in equilibrium and not effected by the interaction with the system. The GKLS has become one of the cornerstones of the theory of open quantum systems.
Solving the Master equation is a difficult
computational problem. The state is described by density matrices
resulting in a computational scaling of at least $O(N^2) $ where $N$ is the size of Hilbert space. Also, in the GKLS equation we address the bath through its ladder operators. Finding them is equivalent to diagonalizing the bath- a very expensive computational task.

To reduce the computational complexity a wavefunction method is desirable. 
The algorithm involves a
stochastic unraveling of wavefunctions.
Stochastic approaches are currently in use in many fields of quantum dynamics such as thermal averaging \cite{gelman2003simulating,ezra2021simulating} and electronic structure methods \cite{baer2013self}.
A stochastic approach has been suggested by Percival and Gisin \cite{gisin1992quantum} for unraveling the GKLS equation.
In their approach the GKLS equation was transformed to stochastic differential equations. This unraveling procedure 
is non unique. 
This has the benefit of the freedom to use the most mathematically convenient choice. The drawback is that the stochastic wavefunction is not associated with a physical description.
An additional problem of this method is that the dynamics are formulated by a non linear differential equation. This increase the difficulty in finding a solution approaching a computational scaling of the size of Hilbert space squared.
A different unraveling approach was developed by Katz, Torrontegui and Kosloff \cite{katz2008stochastic,torrontegui2016activated}. The method partitions the  environment to a primary and secondary bath.
The primary bath termed surrogate Hamiltonian is weakly coupled to the system \cite{baer1997quantum,koch2003surrogate}. Stochastically
the spins of the primary bath are refreshed from the secondary bath.

We now concentrate on a method of unraveling wave functions using stochastic variables modelled on the Collision Model (CM).
The quantum CM, first appeared in  1948 in a paper by Karplus and Schwinger \cite{karplus} followed by work of J. Rau\cite{PhysRev.129.1880}. In the 60s, and later on in the 80s, CMs appeared in works on weak measurements by C.
M. Caves and G. J. Milburn \cite{PhysRevD.33.1643,PhysRevA.36.5543}. In recent years CM became more and more popular due to its simplicity and its ability to be consistent with thermodynamics and the low computational effort in the description of the bath. \cite{kosloff2019quantum}. 
The collision model, models the bath as if it is composed of 
sub units- ancillas, with which the system interact with.
Thus there is one basic assumption in the general CM:
\begin{enumerate}
\item{
The interaction of the system with the bath is described as an interaction between the system with a single ancilla from the Environment.}
\end{enumerate}
We develop a simple collision model with two additional assumptions regarding the bath:
\begin{enumerate}
\item{
Ancillas do not interact with each other.}
\item{
Ancillas are initially uncorrelated.}
\end{enumerate}

This collision model can yield a good description of  real physical example of a system in a diluted gas and also the NV center. 

This general description of CM requires to address two major complexity problems:
\begin{enumerate}
\item{In order to describe the interaction of the system with the ancilla we will have to solve the dynamics of the interaction according to some physical model. This will require to solve the time dependent Schr\"odinger or Liouville equation, a very hard task in general and very expensive for big systems.
Even simpler models to transform only heat between the system and its environment such as resonance between modes of the system to the ancilla particle result in diagonalization of the system's Hamiltonian, which for a system with more then 10 particles is already computationally expensive.}
\item{
The size of the density matrix of a multi particle system grows exponentially with the number of particles and result in huge matrices very fast.}
\end{enumerate}

A significant simplification of the model is obtained by adding the assumption that the interaction between the system and the bath ancilla is much faster than the free dynamics of the the system. This enables us to treat the dissipation and interaction with the bath separately from the internal systems dynamics.

The complexity is further reduced by an implicit treatment of the bath.
We choose to describe the primary system as a system of coupled spins.\\
This representation is employed for three main reasons:
\begin{enumerate}

\item
It is computationally inexpensive.
\item
A system composed of qubits is universal and therefore can simulate other physical systems.
\item 
Such a system can be implemented on a quantum computer.
\end{enumerate}

The collision model creates a Completely Positive Trace Preserving map.  This process leads to a fixed point of the map.
A crucial  tool for the simulation is a partial trace algorithm implemented in a wave functions formalism.

We have developed a stochastic algorithm implemented in wave functions formulation.  The setup  is based on the partial trace implementation. In addition we obtain an intuitive description of the process as an average of partial measurements of the system.
The methods presented in this paper can be implemented in any case of collision model where the environment is composed of qubits as in \cite{li2022steady}, and \cite{o2021stochastic}

\graphicspath{ {./images/} }
\tikzstyle{level 1}=[level distance=25mm, sibling distance=70mm]
\tikzstyle{level 2}=[level distance=25mm, sibling distance=32mm]
\tikzstyle{level 3}=[level distance=45mm, sibling distance=8mm]
\tikzstyle{level 4}=[level distance=45mm, sibling distance=15mm]

\begin{figure}
\begin{tikzpicture}[grow=right,->,>=angle 60]

  \node {$\Ket{\psi}$}
    child {node {$Utr_b\{S_p\Ket{\psi}\otimes\textcolor{red}{\Ket{\beta_{\theta}}}\}$}
        child[dashed] {node{$\textcolor{blue}{(1-p)}\Ket{\psi_{3_2}}$}
            child {node{$\textcolor{blue}{(1-p)}Utr_b\{S_p(\Ket{\psi_{3_2}}\otimes\textcolor{red}{\Ket{\beta_{\theta}}})\}$}}
            child {node{$\textcolor{blue}{(1-p)}Utr_b\{S_p(\Ket{\psi_{3_2}}\otimes\textcolor{red}{\Ket{\beta_{\theta}}})\}$}}
            child {node{$\textcolor{blue}{(1-p)}Utr_b\{S_p(\Ket{\psi_{3_2}}\otimes\textcolor{red}{\Ket{\beta_{\theta}}})\}$}}
            }
        child[dashed] {node{$\textcolor{blue}{p}\Ket{\psi_{3_1}}$}
            child {node{$\textcolor{blue}{(p)}Utr_b\{S_p(\Ket{\psi_{3_1}}\otimes\textcolor{red}{\Ket{\beta_{\theta}}})\}$}
            child[white] {node{$\includegraphics[width=5cm, height=3cm,angle=270]{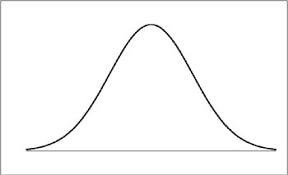}$}}
                child[white]}
            child {node{$\textcolor{blue}{(p)}Utr_b\{S_p(\Ket{\psi_{3_1}}\otimes\textcolor{red}{\Ket{\beta_{\theta}}})\}$}}
            child {node{$\textcolor{blue}{(p)}Utr_b\{S_p(\Ket{\psi_{3_1}}\otimes\textcolor{red}{\Ket{\beta_{\theta}}})\}$}}
            }
      }
    child {node {$Utr_b\{S_p\Ket{\psi}\otimes\textcolor{red}{\Ket{\beta_{\theta}}}\}$}
        child[dashed] {node{$\textcolor{blue}{(1-p)}\Ket{\psi_{2_2}}$}
            child {node{$\textcolor{blue}{(1-p)}Utr_b\{S_p(\Ket{\psi_{2_2}}\otimes\textcolor{red}{\Ket{\beta_{\theta}}})\}$}}
            child {node{$\textcolor{blue}{(1-p)}Utr_b\{S_p(\Ket{\psi_{2_2}}\otimes\textcolor{red}{\Ket{\beta_{\theta}}})\}$}}
            child {node{$\textcolor{blue}{(1-p)}Utr_b\{S_p(\Ket{\psi_{2_2}}\otimes\textcolor{red}{\Ket{\beta_{\theta}}})\}$}}
            }
        child[dashed] {node{$\textcolor{blue}{p}\Ket{\psi_{2_1}}$}
            child {node{$\textcolor{blue}{(p)}Utr_b\{S_p(\Ket{\psi_{2_1}}\otimes\textcolor{red}{\Ket{\beta_{\theta}}})\}$}
            child[white] {node{$\includegraphics[width=5cm, height=3cm,angle=270]{gaussian.jpg}$}}
                child[white]}
            child {node{$\textcolor{blue}{(p)}Utr_b\{S_p(\Ket{\psi_{2_1}}\otimes\textcolor{red}{\Ket{\beta_{\theta}}})\}$}}
            child {node{$\textcolor{blue}{(p)}Utr_b\{S_p(\Ket{\psi_{2_1}}\otimes\textcolor{red}{\Ket{\beta_{\theta}}})\}$}}
            }
        }
    child {node {$Utr_b\{S_p\Ket{\psi}\otimes\textcolor{red}{\Ket{\beta_{\theta}}}\}$}
        child[dashed] {node{$\textcolor{blue}{(1-p)}\Ket{\psi_{1_2}}$}
            child {node{$\textcolor{blue}{(1-p)}Utr_b\{S_p(\Ket{\psi_{1_2}}\otimes\textcolor{red}{\Ket{\beta_{\theta}}})\}$}}
            child {node{$\textcolor{blue}{(1-p)}Utr_b\{S_p(\Ket{\psi_{1_2}}\otimes\textcolor{red}{\Ket{\beta_{\theta}}})\}$}}
            child {node{$\textcolor{blue}{(1-p)}Utr_b\{S_p(\Ket{\psi_{1_2}}\otimes\textcolor{red}{\Ket{\beta_{\theta}}})\}$}}
            }
        child[dashed] {node{$\textcolor{blue}{p}\Ket{\psi_{1_1}}$}
            child {node{$\textcolor{blue}{(p)}Utr_b\{S_p(\Ket{\psi_{1_1}}\otimes\textcolor{red}{\Ket{\beta_{\theta}}})\}$}
            child[white] {node{$\includegraphics[width=5cm, height=3cm,angle=270]{gaussian.jpg}$}}
                child[white]}
            child {node{$\textcolor{blue}{(p)}Utr_b\{S_p(\Ket{\psi_{1_1}}\otimes\textcolor{red}{\Ket{\beta_{\theta}}})\}$}}
            child {node{$\textcolor{blue}{(p)}Utr_b\{S_p(\Ket{\psi_{1_1}}\otimes\textcolor{red}{\Ket{\beta_{\theta}}})\}$}}
            }
      };
\end{tikzpicture}
\caption{
The unraveling tree: The time evolution of the systems wave function $\Ket{\psi}$. Three different interactions with an ancilla $\Ket{\beta_{\theta}}$. Each interaction spawns two wave functions with different wights. This process is repeated with each ancilla interaction.
The Gaussian distribution exhibits that asymptotically the process obeys the central limit theorem.
}
\label{fig:tree}
\end{figure}

\newpage
\graphicspath{ {./images/} }
\tikzstyle{level 1}=[level distance=30mm, sibling distance=72mm]
\tikzstyle{level 2}=[level distance=30mm, sibling distance=65mm]
\tikzstyle{level 3}=[level distance=60mm, sibling distance=35mm]
\begin{figure}
\begin{tikzpicture}[grow=right,->,>=angle 60]

  \node {$\Ket{\psi}$}
    child {node {$Utr_b\{S_p\Ket{\psi}\otimes\textcolor{red}{\Ket{\beta_{\theta}}}\}$}
        child[dashed] {node{$\textcolor{blue}{(1-p)}Utr_b\{S_p(\Ket{\psi_{1}}\otimes\textcolor{red}{\Ket{\beta_{\theta}}})\}$}
            child[dashed]
            {node{$\textcolor{blue}{(1-p)}Utr_b\{S_p(\Ket{\psi_{1_1}}\otimes\textcolor{red}{\Ket{\beta_{\theta}}})\}$}}
            child[dashed]
            {node{$\textcolor{blue}{(p)}Utr_b\{S_p(\Ket{\psi_{1_2}}\otimes\textcolor{red}{\Ket{\beta_{\theta}}})\}$}}
            }
        child[dashed] {node{$\textcolor{blue}{(p)}Utr_b\{S_p(\Ket{\psi_{2}}\otimes\textcolor{red}{\Ket{\beta_{\theta}}})\}$}
            child[dashed]
            {node{$\textcolor{blue}{(1-p)}Utr_b\{S_p(\Ket{\psi_{2_1}}\otimes\textcolor{red}{\Ket{\beta_{\theta}}})\}$}}
            child[dashed]
            {node{$\textcolor{blue}{(p)}Utr_b\{S_p(\Ket{\psi_{2_2}}\otimes\textcolor{red}{\Ket{\beta_{\theta}}})\}$}}
            }
      };
\end{tikzpicture}
\caption{
The reduced unraveling tree: Interaction with a thermal qubit is in agreement with the definitions of Wiener process thus the tree can be reduced to a binary tree form.
}
\label{fig:bintree}
\end{figure}
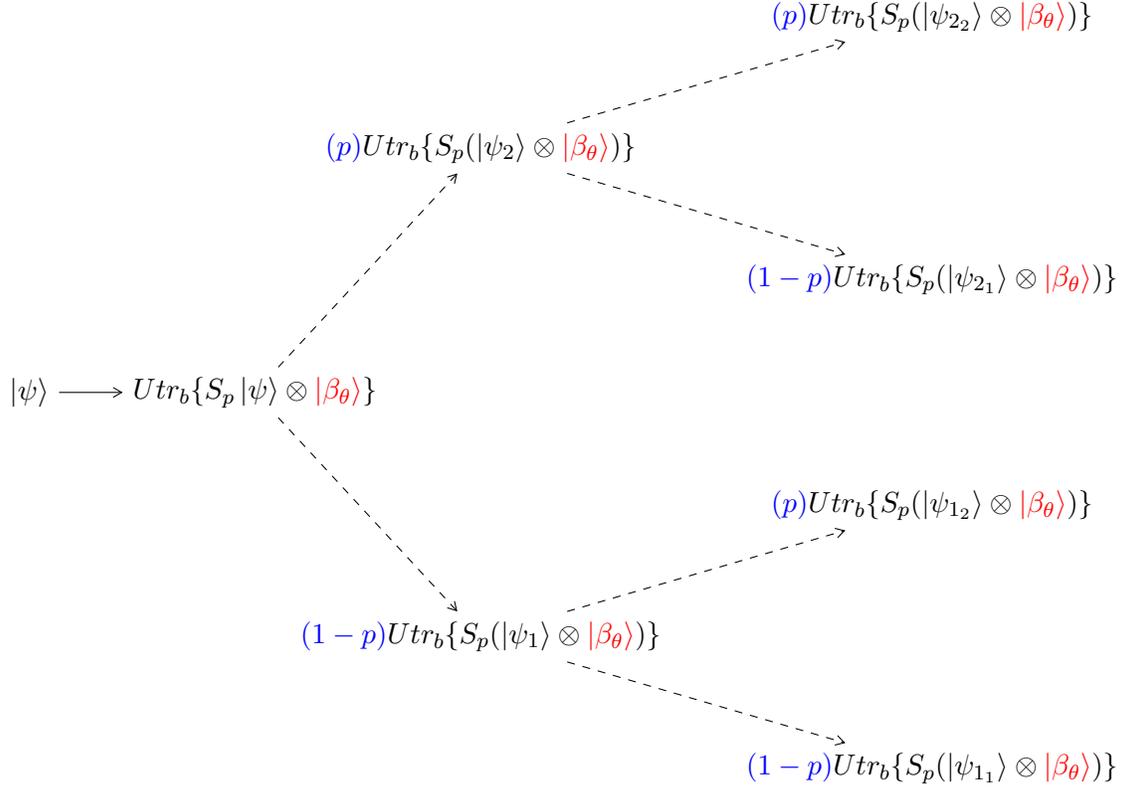
\newpage

\section{\label{sec:Implementation}Implementation}

\subsection{\label{sec:setup}Setup description}
The framework of the derivation assumes a unitary evolution generated by the total Hamiltonian:
\begin{equation}
\hat H=\hat H_S+\hat H_B +\hat H_{SB}~~,   
\label{eq:ghamil}
\end{equation}
composed of the system Hamiltonian $\hat H_S$ environment Hamiltonian $\hat H_B$ and interaction $\hat H_{SB}$.
We assume $\hbar = 1$ all through the paper.

\subsection{\label{sec:spinor}Representation of the spinor}

The model studied is composed of a system of qubits. For such a system the wave function has dimensions of \(1\times 2^N\) when N is the number of qubits. The density matrix representing the system has dimensions of \(2^N\times 2^N\) .
To expand the wave function, we choose a local expansion constructed by the basis of individual components.  
The natural choice for qubits is to construct the wave function in the computational basis as a linear tensor product of the computation base spanning each qubit space.
Each qubit is represented in the computational basis where \(\Ket{0} = \Ket{\downarrow}\) and \(\Ket{1} = \Ket{\uparrow}\).
The complete basis that we will represent the system in is
\begin{equation}
\{\Pi_{i=1}^N\otimes\Ket{\delta_i}\}
\end{equation}

where \(\delta_i = 0,1\)
\begin{equation}
\{\Pi_{i=1}^N\otimes\Ket{\delta_i}\} = \{\Ket{00...0},\Ket{00...1},...,\Ket{01...0},\Ket{01...1} , \Ket{10...0},\Ket{10...1},\Ket{11...0},\Ket{11...1}\}
\end{equation}

The basis is ordered in a raising order of the binary basis.

\subsection{\label{sec:sysbath}System and Baths}

The  multi-qubit systems Hamiltonian \(\hat{H_S}\) is described as: 
\begin{equation}
\hat{H_S} = \sum_{k} \hat{H_k} +\sum_{i,j}\epsilon_{i,j}(\hat \sigma_{+_{i,j}} +\hat \sigma_{-{i,j}})~~~,
\end{equation}
\(\hat{H_k}\) is the free k'th particle Hamiltonian, \(\epsilon_{i,j}\) is the interaction coefficient between the i and j qubits and  is the interaction between the i and j particles in the system.\\
Where N is the number of particles and \[\hat \sigma_{+{i,j}} = \hat I_{2^{i-1}}\otimes\Ket{0}\bra{1}\otimes \hat I_{2^{j-i-1}}\otimes\Ket{1}\bra{0}\otimes \hat I_{2^{N-j}}\]
\[\hat \sigma_{-{i,j}} = \hat I_{2^{i-1}}\otimes\Ket{1}\bra{0}\otimes \hat I_{2^{j-i-1}}\otimes\Ket{0}\bra{1}\otimes \hat I_{2^{N-j}}\]

We represent the state of the system as the density operator \(\hat \rho_S\).\\
The density matrix of the bath \(\rho_{B}\) is composed of uncorrelated ancilla qubits.
\begin{equation}
\hat\rho_{B} = \hat\rho_{b_1} \otimes \hat\rho_{b_2}\otimes...\otimes\hat\rho_{b_n} = \Pi_i \otimes \hat\rho_{b_i} = \Pi_i \otimes \frac{e^{-\beta \hat H_{b_i}}}{Z}
\end{equation}
$\hat H_b$ is the Hamiltonian of the individual ancilla qubit and $\beta$ is the inverse temperature times the Boltzman factor $\beta = \frac{1}{k_B T}$. The fact that the bath is uncorrelated to the system initially is with concent with Kosloff and Dann postulate 2 \cite{dann2021open}.

\subsubsection{\label{sec:level2}Observables}

An observable $\langle \mathcal{O} \rangle$ 
is defined as
\(tr\{\hat{\mathcal{O}}\hat{\rho}\}\).
For a pure state described by a wavefunction  \(\bra{\psi}\hat{\mathcal{O}}\Ket{\psi}\).
In a pure state, it is equivalent to measurement in the wave function formalism.
Let \(\{\Ket{\psi_i}\}\) be an orthonormal basis with \(\Ket{\psi_k} = \Ket{\psi}\) then 
\begin{equation}
    tr\{\hat{\mathcal{O}}\hat{\rho}\} = \sum_{i}^n\bra{\psi_i}\hat{\mathcal{O}}\Ket{\psi}\bra{\psi}\Ket{\psi_i} = 
    \bra{\psi}\hat{\mathcal{O}}\Ket{\psi}
\end{equation}

\subsection{\label{sec:interaction}
System ancilla interaction}

The interaction between the system and the environment is represented as a repeated interaction between the system and a subsystem of the environment. Typically a thermal qubit.
A general unitary interaction is employed.
Therefore the interaction can be expressed by its generator, the interaction Hamiltonian:
\begin{equation}
    \hat U_{int} = e^{-i \hat H_{int}\theta}
\end{equation}
the phase angle $\theta$ has units of time. Since we can write an
exponent as a polynomial sum of $H_{int}$ we get
\begin{equation}
    [\hat U_{int},\hat H_{int}]=0
\end{equation}

\subsection{\label{sec:Dynamics}Dynamics}

In the collision model we assume that the uncorrelated thermal qubit is employed only once. After  interaction   the swapped qubit is traced out. It is therefore assumed that the state of the bath is unchanged. 
This assumption is in accordance of an uncorrelated infinitely large bath  and imposes that the systems state is completely dependent on its previous state.
The last remark is a definition of Markovianity and is in accordance with postulate $ 4$ of Kosloff and Dann \cite{dann2021open}
\begin{equation}
\Lambda_{t} = \Lambda_{t-s}\Lambda_{s}
\end{equation}
The dynamical map propagates density operators. A reduced map
generated by a global  Hamiltonian Eq. (\ref{eq:ghamil}), from an  initial uncorrelated state defines a Kraus map \cite{kraus1974operations}. Such a map $\Lambda$ is a Completely Positive Trace Preserving (CPTP) map on the system  \cite{dann2021open}. 
The generator of the dynamics  is defined as:
\begin{equation}
\mathcal{L} = \lim_{t \to 0} \frac{\Lambda(t) - \hat{\mathcal{I}}}{dt}
\end{equation}
Under the assumption that the collision period is much shorter than the interval between  collisions we can write the generator as follow:
\begin{equation}
\label{eq:poisson}
\mathcal{L} = -i[\hat H,\bullet]+\gamma(tr_b\{\hat U_{int}\bullet \otimes \hat \rho_{b}\hat U_{int}^{\dagger}\}-\hat{\mathcal{I}}\bullet)~~~, 
\end{equation}
where \(\gamma\) is the collision rate,
eq. (\ref{eq:poisson}).
The reduced description  will have the form
\begin{equation}
\frac{d}{dt}\hat \rho_S = \mathcal{L}(\hat \rho_S) = -i[\hat H_{S},\hat \rho_S] +\gamma(tr_b\{\hat U_{int} \hat \rho_S \otimes \hat \rho_b \hat S_{p}^{\dagger}\}-\hat \rho_S)
\end{equation}

This structure has the Poissonian GKLS form   \cite{lindblad1976generators}. 
Assuming the swap is instantaneous we can write the integrated form of the dynamics as a sequence of collision events, where $\gamma$ is determined by the average propagation time dt
A single collision event can be described by the super operator $\mathcal{M}$ acting on the density operator $\rho_s$
\begin{equation}
    \mathcal{M}(\hat H_S,\gamma,\theta,\beta)\hat \rho_S=(\hat U(\hat H_S,dt))tr_b\{\hat U_{int}(\theta)\hat \rho_S \otimes \hat  \rho_{b} \hat U_{int}(\theta)^{\dagger})\}(\hat U^{\dagger}(\hat H_S,dt))
\end{equation}
And $k$ consecutive collisions
\begin{equation}
\label{dynamics}
    \prod_{i=0}^k\mathcal{M}_i(\hat H_S,\gamma,\theta,\beta)
\end{equation}

\subsection{\label{DM} Unraveling of the density operator }

The density operator $\hat \rho$ completely describes the state of the quantum system. Any observable is determined by the relation $\langle \mathcal{O} \rangle = tr \{ \hat \rho \mathcal{O} \}$.  
The density operator was introduced by von Neuman to describe  statistical 
phenomena in quantum mechanics \cite{von2018mathematical}. It was observed that  a pure state in an entangled system described by a wave function is reduced to a mixed state when observing the state of a subsystem. 

The statistical character of the density operator is reflected by the unraveling 
to an average of outer product of wave functions:
\begin{equation}
    \hat \rho = \sum_k p_k | \psi_k \rangle \langle \psi_k |
\end{equation}
where $\{\Ket{\psi_k}\}$ is the set of unraveling wavefunctions (not necessarily orthogonal).
The unraveling set $\{\ket{\psi}\}$ is not unique which allows freedom which we will exploit.
A straight forward unraveling is obtained by
diagonalizing the density operator. 

\subsection{\label{sec:StochasticU}Stochastic unravelling}

The present study employs a stochastic unravelling  scheme based on the flowing lemma:
let \(\theta\) be a random phase wavefunction then
\begin{equation}
    \lim_{N \to \infty}\frac{1}{N}\sum_{j,k=1}^N e^{i(\theta_j-\theta_k)} = \delta_{j,k}
\end{equation}

Let \(\ket{\psi_{\theta}}\) be a wave function composed of an equal superposition of an arbitrary orthonormal basis $\{\Ket{n}\}$ of size N with random phase $\theta$
\begin{equation}
    \Ket{\psi_{\theta}} = \frac{1}{\sqrt{N}}\sum_{n=0}^{N-1}e^{i\theta_n}\Ket{n}
\end{equation}
The identity operator $\hat I_N $ can be resolved by an infinite sum of random wave functions.
\begin{equation}
    \lim_{K \to \infty} \frac{N}{K} \sum_{i=1}^K \Ket{\psi_{\theta_i}}\bra{\psi_{\theta_i}}= \hat I_N
\end{equation}
Since every density matrix can be diagonalized $\hat{\rho}$ can be decomposed to:
\begin{equation}
    \hat{\rho} = \sum_{n=1}^N p_i\Ket{n}\bra{n}
\end{equation}
This allow to unravel the density operator with stochastic wave functions. We will prove, that for every basis that an arbitrary $\hat \rho $ is diagonal in we can create a set of stochastic wave functions that unravel $\hat \rho $.
\begin{equation}
\label{eq:ranf}
    \Ket{\phi} = \sqrt{N}\sum_{n=1}^N\sqrt{p_n}\ket{n}\bra{n}\ket{\psi_{\theta}}
\end{equation}
\(\Ket{\phi}\) is a normalized wave function:
\begin{equation}
    \braket{\phi}= \sqrt{N}\sum_{n=1}^N\sqrt{p_n}\bra{\psi_{\theta}}\ket{n}\bra{n}\sqrt{N}\sum_{n=1}^N\sqrt{p_n}\ket{n}\bra{n}\ket{\psi_{\theta}}= 1
\end{equation}
The density operator $\hat \rho$ converges to an average over the outer product of $\Ket{\phi}$ under the condition that it is a stochastic wave function.
\begin{equation}
   \hat \rho = \sum_{n=1}^N\sqrt{p_n}\ket{n}\bra{n}\sum_{n=1}^N\sqrt{p_n}\ket{n}\bra{n} = \sum_{n=1}^N\sqrt{p_n}\ket{n}\bra{n}~\hat I~\sum_{n=1}^N\sqrt{p_n}\ket{n}\bra{n}=
    \nonumber
\end{equation}
\begin{equation}
\nonumber
    \sum_{n=0}^{N-1}\sqrt{p_n}\ket{n}\bra{n}\lim_{K \to \infty} \frac{N}{K} \sum_{i=1}^K \Ket{\psi_{\theta_i}}\bra{\psi_{\theta_i}}\sum_{n=0}^{N-1}\sqrt{p_n}\ket{n}\bra{n}=
\end{equation}
\begin{equation}
    \lim_{K \to \infty} \sum_{i=1}^K\frac{1}{K}\sqrt{N}\sum_{n=0}^{N-1}\sqrt{p_n}\ket{n}\bra{n}\Ket{\psi_{\theta_i}} \sqrt{N}\sum_{n=0}^{N-1}\sqrt{p_n}\bra{\psi_{\theta_i}}\ket{n}\bra{n} =
\end{equation}
\begin{equation*}
    \lim_{K \to \infty} \frac{1}{K}\sum_{i=1}^K\Ket{\phi_i}\bra{\phi_i} \;\;\;\;\;\square
\end{equation*}

A single wavefunction is a pure state and therefore cannot
describe a statistical distribution. 
In particular  a thermal state is never pure.
To overcome this issue the stochastic unraveling method is employed to represent the thermal ancilla qubits that collide with the system in accordance with Eq. (\ref{eq:ranf}): 
\begin{equation}
    \hat \rho = \sum_{j=0}^{N-1} p_j \Ket{\psi_j}\bra{\psi_j}
\end{equation}

\begin{equation}
\Ket{\beta_j} = \sum_{i=0}^1\sqrt{N\frac{e^{-\beta \omega_i}}{Z}}\Ket{\omega_i}\bra{\omega_i}\Ket{\psi_{\theta_j}} =\sum_{i=0}^1\frac{\sqrt{N}e^{-\frac{\beta}{2}\omega_i}}{\sqrt{Z}}\Ket{\omega_i}\bra{\omega_i}\Ket{\psi_{\theta_j}}  \end{equation}
\begin{eqnarray}
\hat \rho_b=\lim_{K\to \infty}\frac{1}{K}\sum_{j=1}^{K}\ket{\beta_j}\bra{\beta_j} =
    \frac{e^{-\beta \hat H}}{Z}
\end{eqnarray}

An ensamble average of interactions between the system and many \(\Ket{\beta_j}\) will converge to an interaction between the system and a thermal qubit 
\begin{equation}
   \lim_{K\rightarrow\infty}\frac{1}{K}\sum_{j=1}^K \hat U_{int}\Ket{\psi}\otimes\Ket{\beta_j}\bra{\psi}\otimes\bra{\beta_j}\hat U_{int}^{\dagger} = 
\end{equation}
\begin{equation*}
    \hat U_{int}\Ket{\psi}\bra{\psi}\otimes\Big(\lim_{K\rightarrow\infty}\frac{1}{K}\sum_{j=1}^n\Ket{\beta_j}\bra{\beta_j}\Big)\hat U_{int}^{\dagger} = \hat U_{int}\hat \rho_s\otimes \hat \rho_b \hat U_{int}^{\dagger}
\end{equation*}
This method enables  to describe a swap interaction with a thermal qubit in the language of wave functions. In order to completely restore Eq. (\ref{dynamics}) we need to translate the partial trace operation into wave function terminology.

\subsection{\label{sec:Partrace} Partial trace in a wave function description}

Partial trace is an essential operation in obtaining the state of a subsystem from a composite state. When the subsystem is entangled with its complementary system
the partial trace operation will lead to a mixed state.
 
To represent such a state with a wave function, stochastic unraveling will be employed. The algorithm is designed to include in the wave functions, the correct probabilities such that an average of their outer product will reproduce the reduced subsystem state.

Even though the algorithm presented can be generalized for the tracing out of any number of qubits, in this paper we present the algorithm of tracing out a single qubit.
Specifically we assume a system of $n$ spins and we trace out the \(k_{th}\) spin.
\(for\;\; b\in [0,2^{k-1}]\;\; ,\;\; a\in [0,2^{n-k}]\) and $i\in [0,1]$

We define \(\Ket{\phi}^i\)
\begin{equation}
\label{ptrace}
    \Ket{\phi}^i_{b*2^{n-k}+a} = \frac{\Ket{\psi}_{(2b+i)*2^{n-k}+a}}{N_i}
\end{equation}
where N is the normalization factor and also the square root of the classical probability of this state.
\newline
\(\Ket{\phi}^i\) is not a random choice, it is embedding the physical meaning of the measurement of the environment.
\newline
\(\Ket{\phi}^0\) is a normalized vector of all the elements in \(\Ket{\psi}\) condition on the state of traced out particle  \(\Ket{0}\).
\newline
\(\Ket{\phi}^1\) is a normalized vector of all the elements in \(\Ket{\psi}\) condition on the state of the traced out particle  \(\Ket{1}\).
\newline
The main result is 
\begin{equation}
\label{eq:partrace}
     tr_{k}\{\Ket{\psi}\bra{\psi}\} =  \sum_{i=0}^1 N_i^2\Ket{\phi}^i\bra{\phi}^i
\end{equation}

The proof is described in appendix (\ref{partialtrace}).

The result above, beside complexity reduction also underlines the measurement postulate in quantum mechanics and the equivalence of the partial trace with partial measurement. The partial trace is a sum of the system's possible states after the traced out particle has "collapsed" in to its possible states with the adequate probability.

\subsection{\label{sec:stochpartrace} Stochastic Partial trace}
The main result of section [\ref{sec:Partrace}] is Eq. (\ref{eq:partrace}), giving an unravelling of the mixed state. The description of the process requires us to operate on each wave function of the unraveled tree separately as in figure \ref{fig:tree}. As will be described in the next section, the probabilistic nature of the mixed state enable to employ a Monte Carlo algorithm to randomly chose only one of the functions. we have developed a generalized Monte Carlo algorithm for tracing out more then one qubit which is not presented here.
\begin{equation}
\label{eq:Monte}
    tr_{b_{x_r}}{\Ket{\psi}} =
        \begin{cases}
            \Ket{\phi}^0 \quad \text{if} \quad  x_r < N_0^2 \\ \Ket{\phi}^1 \quad  \text{if}\quad x_r \ge N_0^2
        \end{cases}
\end{equation}
The Monte Carlo algorithm induces the probability  $ N_0^2\quad is \quad tr_{b_{x_r}}{\Ket{\psi}} = \Ket{\phi}^0$ and the complementary probability  $N_1^2 \quad is \quad tr_{b_{x_r}}{\Ket{\psi}} = \Ket{\phi}^1$.

\subsection{\label{sec:process} Describing the branching process.}

The branching process is a sequence of free dynamics following a unitary collision
Cf. Sec. \ref{sec:Dynamics}.
This process is described in the language of wave functions, allowing an efficient algorithm by wave functions, that its unravelling is converging to the density operator dynamics representation. 
By combining the stochastic unravelling and partial trace  presented in (\ref{eq:ranf}) and (\ref{ptrace}) we can construct Eq. (\ref{dynamics}) that represent the consecutive collisions of the density matrix with a thermal particle :

In details, every wave function \(\Ket{\psi}\)
will undergo three consecutive operations:
\begin{enumerate}
\item{Interaction with a thermal wave function- $\hat U_{int}(\Ket{\psi}\otimes\Ket{\beta})$.}
\item{Stochastic partial trace - ${tr_{N+1_{x_r}}}\{ \hat U_{int}\Ket{\psi}\otimes\Ket{\beta}\}$.}
\item{Free dynamic of the system- 
${\hat U} tr_{N+1_{x_r}}\{ \hat U_{int}\Ket{\psi}\otimes\Ket{\beta } \} $.
}
\end{enumerate}

To accurately restore Eq. (\ref{eq:wave1}), $\Ket{\psi}$ will accumulate $k$ thermal wave functions $\Ket{\beta}$ $(k\rightarrow \infty)$. The partial trace will yield two different wave functions with different probabilities, thus we will have to average the outer product of all the outcomes with the correct weighs.
This procedure corresponds to a single collision.

\begin{equation*}
    \lim_{K \to \infty}\frac{1}{K}\sum_{j=1}^K\int_{x_r=0}^1 \hat U tr_{{N+1}_{x_r}}\{ \hat U_{int} \Ket{\psi}\otimes\Ket{\beta_j}\}tr_{{N+1}_{x_r}}\{\bra{\psi}\otimes\bra{\beta_j} \hat U_{int}^{\dagger} \}\hat U^{\dagger}dx_r
\end{equation*}

\begin{equation*}
    \lim_{K \to \infty} \hat U tr_{N+1}\{ \hat U_{int} \Ket{\psi}\bra{\psi}\otimes \frac{1}{K}\sum_{j=1}^K\Ket{\beta_j}\bra{\beta_j} \hat U_{int} \}\hat U^{\dagger}
\end{equation*}

\begin{equation}
\label{eq:wave1}
     \hat U (tr_{N+1}\{\hat U_{int} (\hat \rho_s \otimes \hat \rho_B)\hat U_{int}^{\dagger} \})\hat U^{\dagger}
\end{equation}

For repeated collisions this process will recur for every outcome as shown in figure (\ref{fig:tree}).
Eq. (\ref{dynamics}) is restored by consecutively employing Eq. (\ref{eq:wave1}) n times.
For a mathematical description we will define the super operator
\begin{equation}
\hat{\mathcal{G}} (\psi,dt,\phi,x_r,\beta) = \hat U tr_{{N+1}_{x_r}}\{ \hat U_{int} \Ket{\psi}\otimes\Ket{\beta_j}\}tr_{{N+1}_{x_r}}
\end{equation}

\begin{equation}
\label{eq:tree}
    \lim_{K \to \infty}\frac{1}{K}\sum_{j=1}^K \int_{x_r=0}^1 \hat{\mathcal{G}}^n(\psi,dt,
    \phi,x_r,\beta) \hat{\mathcal{G}}^{\dagger n} (\psi,dt,\phi,x_r,\beta)dx_r = \hat \rho_n
\end{equation}

\subsection{\label{sec:converge}
Stochastic convergence}

This process is extremely computationally expensive, since it grows exponentially with each collision as $O(2^{nk})$ where n is the number of collisions and k is the number of thermal wave functions interacting with each possible state of the system. This resolve in a branching tree illustrated in Fig. \ref{fig:tree}.

A  solution  for this problem is to exploit the stochastic nature of the process in three ways.
\begin{enumerate}
\item{Reduction to a binary tree due to a Wiener process.
If we look at the path of a single wave function (single branch in the tree in Fig. \ref{fig:tree}) we observe that the systems wave function interacts in every collision with a stochastic thermal wave function. The process satisfies the conditions of a Weiner process - the process has a fixed initial condition and the stochastic part of the bath particle in every collision has a mean 0 and a variance $\sigma^2$.
Therefore for sufficiently long  process, where each collision $U_{int}(\Ket{\psi_k}\bra{\psi_k}\otimes \rho_b)U_{int}^{\dagger}$ is represented as the average of the outer product of K $U_{int}(\Ket{\psi_k}\otimes \Ket{\beta_{j}})$ and thus splited into K  branches, can be represented by only one $\Ket{\beta_j}$. For many collision it will,  follow the Weiner process, undergo a Brownian motion and the process will converge to a consecutive interaction with a thermal qubit. As a result the unravelling of the tree in figure \ref{fig:bintree} will converge to the unravelling of the larger tree in figure \ref{fig:tree}.

As can be seen in figure \ref{fig:bintree} this method will result in a properly weighed sample of the binomial distribution around the most probable state.}
\item{
The nature of tracing out a single bath qubit results in a mixed state composed of two pure states eq. (\ref{eq:partrace}).
Using the property of Weiner process, computation of all possibilities with the correct weights will be resolved in a binary tree with changing probability weights. 
We used a Monte Carlo stochastic partial trace algorithm sec. \ref{sec:Partrace} in order to stochastically chose one of the two wave functions constructing the mixed state imposed by the partial trace in every step.}
\item{
Based on the central Limit Theorem, the average of a sequence of independent and identically distributed random variables drawn from a distribution of expected value given by $\mu$ and finite variance given by $\sigma^2$ will converge in probability to a normal distribution.
The multidimensional Central Limit Theorem generalize the theory and state that a random vector (satisfying the vector space axioms) will converge in probability to a multi-variant Gaussian. Mathematically
\begin{equation}
    \sqrt{n} (\hat{X_n}-\mu) \rightarrow^d \mathcal{N}(0,\Sigma)
\end{equation}
where $\Sigma$ is the covariance matrix.
}
\end{enumerate}

Employing Eq. (\ref{eq:tree}) the average of the outer product of all wave function possibilities with the adequate probabilities represented by the tree in Fig. \ref{fig:tree} converges to the density matrix $\rho_n$ satisfying the collision model in sec. \ref{sec:Dynamics}.
Thus a sample of the outer product of $K$ identical wave functions undergoing n consecutive collisions by eq. (\ref{eq:tree}):
$\mathcal{G}^n(\psi,dt,\phi,x_r,\beta){\mathcal{G}^n}^\dagger(\psi,dt,\phi,x_r,\beta)$
is a sequence of independent and random variables drawn from a distribution of expected value $\rho_n$ and a finite variance and thus
\begin{equation}
   \lim_{K \to \infty} \frac{1}{K}\sum_{i=1}^K\mathcal{G}^n(\psi,dt,\phi,x_r,\beta){\mathcal{G}^n}^\dagger(\psi,dt,\phi,x_r,\beta) =  \hat \rho_n
\end{equation}
Moreover with n- number of collisions increasing we expect  convergence in probability
\begin{equation}
    \frac{1}{K}\sum_{i=1}^K\mathcal{G}^n(\psi,dt,\phi,x_r,\beta){\mathcal{G}^n}^\dagger(\psi,dt,\phi,x_r,\beta) - \hat \rho_n \rightarrow^d
\mathcal{N}(0,\frac{\Sigma}{K})
\end{equation}
Central limit theorem redundant mathematical use of Brownian motion. Yet, we have found it useful for physical intuition of the process.

\section{\label{sec:Results}Results}
\subsection{\label{sec:convergence}Convergence}

To demonstrate the approach in accordance with sec. \ref{sec:converge}, we will study a specific example of a unitary interaction.  A partial swap between the last particle in a system of qubits and an uncorrelated thermal qubit as interaction is specifically chosen.
This type of interaction has been addressed in the collision model review by Ciccarello, Lorenzo, Giovannetti and Palma \cite{CICCARELLO20221}.

In a two qubit system the swap algorithm becomes the swap gate:
\begin{equation}
\hat S = \Ket{0}\bra{0}\otimes\Ket{0}\bra{0}+\Ket{1}\bra{1}\otimes\Ket{1}\bra{1}+\Ket{1}\bra{0}\otimes\Ket{0}\bra{1}+ \Ket{0}\bra{1}\otimes\Ket{1}\bra{0}=
    \begin{pmatrix}
    1&0&0&0\\
    0&0&1&0\\
    0&1&0&0\\
    0&0&0&1
    \end{pmatrix}
\end{equation}
In a bigger system composed of $N$ qubits, an operation that swaps between the $i~th$ and $j~th$ qubits the swap algorithm looks similar:

let \begin{equation}\prod_{k=0}^{i-1}\otimes   I_{2^k}\otimes\Ket{\delta}\otimes\prod_{k=i+1}^{N}\otimes I_{2^k} = \Ket{\delta_i}\end{equation}
\begin{equation}\prod_{k=0}^{i-1}\otimes   I_{2^k}\otimes\bra{\delta}\otimes\prod_{k=i+1}^{N}\otimes I_{2^k} = \bra{\delta_i}\end{equation}
A swap operation between particles i and j can be described as
\begin{equation}
\hat S_{i,j} = \Ket{0_i}\bra{0_i}\Ket{0_j}\bra{0_j}+\Ket{1_i}\bra{1_i}\Ket{1_j}\bra{1_j}+\Ket{0_i}\bra{1_i}\Ket{1_j}\bra{0_j}+\Ket{1_i}\bra{0_i}\Ket{0_j}\bra{1_j}
\end{equation}

Swap is a unitary operation - see appendix (\ref{ap:unitaryS})
\begin{equation}
    \hat S \hat S^{\dagger} =  \hat S_{k,m} \hat S_{k,m}^{\dagger} = \hat I_{2^{n}} 
\end{equation}
$\hat S$ is unitary and real, therefor \(\hat S = \hat S^{\dagger}\).

The unitarity of the swap operator makes it a preferable candidate to simulate interaction.
It has the advantage that it can be realized on a quantum computer.

The partial swap is defined as:
\begin{equation}
    \hat S_p = \cos(\theta)\hat I_{2^n}+i \sin (\theta) \hat S
\end{equation}
The partial swap, like the full swap, is also a unitary operation $\hat S_p \hat S_p^{\dagger} = \hat I$- see appendix (\ref{ap:unitarySp}). Unlike the full swap, the partial swap induces a spectrum of interaction strengths and maintains correlation between all particles in the system after reduced description of the setup as we will see in Sec. (\ref{sec:interaction}).
For this reason we chose it as a preferable candidate for the interaction between the ancilla and the system.
\newline
Since \(S_p\) is unitary. by definition we can write 
\begin{equation}
\hat S_p = e^{-i \hat H_{int}\theta}
\end{equation}
\(\theta\) can be either a phase, angle of interaction or time of interaction. 
Since we can write an exponent as a polynomial sum of \(\hat H_{int}\) we get
\begin{equation}
[\hat S_p,\hat H_{int}]=0
\end{equation}

if we call the unravelling of K processes
\begin{equation}
    \frac{1}{K}\sum^K \mathcal{G}^n{\mathcal{G}^n}^{\dagger} = \Theta_n(K)
\end{equation}
We expect the distance 
\begin{equation}
    |\Theta_n(K) - \rho_n|
\end{equation}
To converge in probability to $\mathcal{N}(0,\frac{\Sigma}{K})$- a normal distribution around zero.
In order to examine the convergence rate of our model, we chose to observe the variance of the distribution of this distance and expect it to converge as $\frac{1}{K}$.
The distance function we chose was the variance between each element of the unravelling and the density matrix $\rho_s$ of the system under going the same dynamic in the density matrix form- Eq. (\ref{dynamics}).
$N^2$ being the dimension of the system, K- the number of realizations and n- the number of collisions.
\begin{equation}
    D(\mathcal{G},\rho,n,N,K) = \frac{1}{N^2}\sum_{i,j=1}^{N^2}|\rho_{n_{i,j}}-\Theta(K)_{n_{i,j}}|^2
\end{equation}
\begin{figure}
\includegraphics{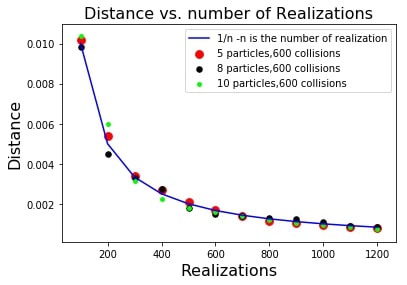}
\caption{
The distance function defined by a normalized sum of the absolute value squared of the euclidean distance between same elements of $\rho_n$ and $\Theta_n$. That is taken as a normalization of $\mathcal{D}(\mathcal{G},\rho,n,N) = \frac{1}{N^2}\sum_{i=1}^{N^2}|\rho_{n_{i,j}}-\Theta_{n_{i,j}}|^2$. We expect that $\cal{D}$ will converge as $\frac{1}{K}$. Since we expect the variance to converge as $\frac{1}{K}$, we expect the variance of each element also to converge as $\frac{1}{K}$ and therefore that $\mathcal{D}$, that is a sum of variance element of each matrix element will converge as $\frac{1}{K}$ as well. The graph exhibits the ${\cal D}$ function for 3 systems undergoing 600 collisions. We can see a decay to zero as $\frac{1}{K}$ for a system composed of 5,8 and 10 particles as expected
}
\label{fig:distance}
\end{figure}
According to the multidimensional Central Limit Theorem we expect a convergence of the covariance matrix as $\frac{1}{K}$, and thus the variance of every element $|\Theta_{i,j}-\rho_{i,j}|$ to $\sigma_{i,j}^2$. Since each element of $\Theta$ has the same dependency on the stochastic variable we expect $\sigma_{i,j}=\sigma_{k,l}$.
Thus we expect D to approximate the average of all element variances.
\begin{equation}
    D(\mathcal{G},\rho,n,N,K) = \frac{1}{N^2}\sum_{i,j=1}^{N^2}|\rho_{n_{i,j}}-\Theta_{n_{i,j}}(K)|^2 \approx  \sum_{i,j}^{N^2}\frac{1}{K}\sum_{k}^K|\rho_{n_{i,j}}-\Theta_{n_{i,j_k}}|^2
\end{equation}
From the central limit theorem we expect that the sum of all element variances will decay as $\frac{1}{K}$ with K the number of realization
\begin{equation}
    \sum_{i,j}^{N^2}\frac{1}{K}\sum_{k}^K|\rho_{n_{i,j_k}}-\Theta_{n_{i,j_k}}|^2 \sim \frac{1}{K}
\end{equation}
thus
\begin{equation}
    D(\mathcal{G},\rho,n,N,K) \sim \frac{1}{K}
\end{equation}
As we can seen in figure \ref{fig:distance} we observe that
${\cal D}$ indeed converges as expected.

\section{
\label{conclusions}
Conclusions}

In this paper we developed the basic tools for unraveling open system
dynamics with wavefunctions
based on a collision model.
The stochastic averaging of these wavefunctions converges 
for all expectation values,
equivalent to the  density operator formalism. The basic algorithm can be divided into three steps:
\begin{enumerate}
    \item  Implementing a Monte Carlo stochastic partial trace algorithm. Sec. \ref{sec:Partrace},\ref{sec:stochpartrace} .
    \item Restoring a mixed state by unraveling of stochastic wave functions. This was achieved for the bath particle colliding with the system and for the system itself after averaging over many stochastic system wave functions undergoing the dynamics. Sec. (\ref{sec:StochasticU})
    \item Using statistical properties such as the central limit theorem we observed convergence to the density matrix described dynamics. The convergence was achieved with a relatively small number of realizations. This property results in high computation efficiency.
\end{enumerate}
An illustration of the 
algorithm  is carried out in Sec. (\ref{sec:convergence}). 

The algorithm developed contains  a nonlinear component. In
the implementation of the Monte Carlo algorithm in Sec. (\ref{sec:stochpartrace}), one of the wave functions composing the mixed state is selected by partial trace. The probability to choose each of the states, is calculated from the normalization factor of one of the states. This requires to first calculate one  possible outcome. This outcome   might not be used in the next step. If the selected wavefunction is the one that was not calculated, the other wave function has to be recalculated. Thus, the non-linearity results from renormalizing the wavefunction and the possibility  of computing an additional wavefunctions if it was absent in the Monte Carlo lottery.

The modeling method addresses a major problem of the cost of simulating open quantum system. The wavefunction representation
reduces the memory requirement and as the system becomes larger the speed of convergence to the
full simulation also increases. We therefore expect a reduction in computational cost of up to a factor of $N$ where $N$ is the size of Hilbert space.
As a result the boundaries of possible simulations are stretched.

In addition the wavefunction method allows additional insight in the physical process taking place from the viewpoint of a single event.

Finally the simulation can be implemented on quantum computers- Due to the fact that all operations are unitary and the description of the setup is by wavefunctions.


\begin{acknowledgments}

We thank Christiane Koch, Gil Katz, and  Florian Habecker for  sharing their insight.
Work supported by the Israel Science Foundation (Grants No. 510/17 and 526/21).

\end{acknowledgments}
\newpage
\appendix

\section{\label{appendix}Appendixes}
\subsection{\label{ap:unitaryS}Proof of Unitarity of $\hat S$}
\begin{equation}
\hat S_{i,j} \hat S_{i,j}^{\dagger} = (\Ket{0_i}\Ket{0_j}\bra{0_j}\bra{0_i}+\Ket{0_i}\Ket{1_j}\bra{1_j}\bra{0_i}+\Ket{1_i}\Ket{0_j}\bra{0_j}\bra{1_i}+\Ket{1_i}\Ket{1_j}\bra{1_j}\bra{1_i}( \end{equation}
\begin{equation*}
(\Ket{0_j}\Ket{0_i}\bra{0_i}\bra{0_j}+\Ket{0_j}\Ket{1_i}\bra{1_i}\bra{0_j}+\Ket{1_j}\Ket{0_i}\bra{0_i}\bra{1_j}+\Ket{1_j}\Ket{1_i}\bra{1_i}\bra{1_j}) = 
\end{equation*}
\begin{equation*}
    \Ket{0_i}\Ket{0_j}\bra{0_j}\bra{0_i}\Ket{0_j}\Ket{0_i}\bra{0_i}\bra{0_j}+0+0+0+
\end{equation*}
\begin{equation*}0+0+\Ket{0_i}\Ket{1_j}\Ket{1_j}\Ket{0_i}\bra{0_i}\bra{1_j}+0+\end{equation*}\begin{equation*}0+\Ket{1_i}\Ket{0_j}\bra{0_j}\bra{1_i}\Ket{0_j}\Ket{1_i}\bra{1_i}\bra{0_j}+0+0+\end{equation*}\begin{equation*} 0+\Ket{1_i}\Ket{1_j}\bra{1_j}\bra{1_i}\Ket{1_j}\Ket{1_i}\bra{1_i}\bra{1_j}) = 
\end{equation*}\begin{equation*}
    \Ket{0_i}\Ket{0_j}\bra{0_i}\bra{0_j} +\Ket{0_i}\Ket{1_j}\bra{0_i}\bra{1_j}+\Ket{1_i}\Ket{0_j}\bra{1_i}\bra{0_j}+\Ket{1_i}\Ket{1_j}\bra{1_i}\bra{1_j}=
\end{equation*}\begin{equation*}
    (\Ket{0_i}\bra{0_i}+\Ket{1_i}\bra{1_i})(\Ket{0_j}\bra{0_j}+\Ket{1_j}\bra{1_j})=\hat I_{2^N}
\end{equation*}
\subsection{\label{ap:unitarySp}Proof that the partial swap is unitary.}
\begin{equation}
    \hat S_p = \cos(\theta) \hat I_{2^n}+i \sin (\theta) \hat S
\end{equation}
\begin{multline}
\hat S_p \hat S_p^{\dagger} = (cos(\theta)\hat I_{2^n}+i \sin(\theta) \hat S)(\cos(\theta)\hat \hat I_{2^n}+i\sin(\theta) \hat S)^{\dagger} =\\
    \cos^2(\theta)\hat I_{2^n}*\hat I_{2^n}+\sin^2(\theta)\hat S \hat S^{\dagger}+i\sin(\theta)\cos(\theta)\hat I_{2^n}\hat S^{\dagger}-i\sin(\theta)\cos(\theta)\hat I_{2^n}=\\
    \hat I_{2^n}(\cos^2(\theta)+\sin^2(\theta))+\hat S(i\sin(\theta)\cos(\theta)-i\sin(\theta)\cos(\theta)) =
    \hat I_{2^n}+0 = \hat I_{2^n}    
\end{multline}

\subsection{\label{partracenp} Tracing out the n'th qubit tensor thermal qubit} 
\begin{equation}
    tr_{n+1}\{\hat S_{n,n+1} (\Ket{\psi}\bra{\psi}\otimes\ket{\beta}\bra{\beta})\hat S_{n,n+1}^{\dagger}\} =  tr_{n}\{\ket{\psi}\bra{\psi}\}\otimes\ket{\beta}\bra{\beta}
\end{equation}

\begin{equation}
    \hat I_{2^{n-1}}\otimes \Ket{0} = \Ket{\mathcal{O}} \;\;\; \hat I_{2^{n-1}}\otimes \Ket{1} = \Ket{\mathcal{I}}
\end{equation}

\begin{equation}
    \hat S_{n,n+1} = \hat I_{2^{n-1}}\otimes(\Big(\sum_{i=0}^1 \Ket{i}\bra{i}\otimes \Ket{i}\bra{i} + \sum_{k\neq j=0}^1 \Ket{k}\bra{j}\otimes \Ket{j}\bra{k}\Big)
\end{equation}
\begin{equation}
    tr_{n+1}\{\hat S_{n,n+1} (\Ket{\psi}\bra{\psi}\otimes\ket{\beta}\bra{\beta}) \hat S_{n,n+1}^{\dagger}\} =
\end{equation}
\begin{multline*}
    \sum_{l=0}^1 I_{2^{n}}\otimes \bra{l} \Big( I_{2^{n-1}}\otimes \Big(\sum_{i=0}^1 \Ket{i}\bra{i}\otimes \Ket{i}\bra{i} + \sum_{k\neq j=0}^1 \Ket{k}\bra{j}\otimes \Ket{j}\bra{k}\Big) \ket{\psi}\bra{\psi}\otimes
\ket{\beta}\bra{\beta}\\
\Big( I_{2^{n-1}}\otimes \Big(\sum_{i=0}^1 \Ket{i}\bra{i}\otimes \Ket{i}\bra{i} + \sum_{k\neq j=0}^1 \Ket{k}\bra{j}\otimes \Ket{j}\bra{k}\Big) =\\
\end{multline*}
\begin{multline*}
I_{2^{n}}\otimes \bra{0} \Big(\Ket{\mathcal{O}
 }\bra{\mathcal{O}
 }\otimes \Ket{0}\bra{0} + \Ket{\mathcal{I}}\bra{\mathcal{O}}\otimes \Ket{0}\bra{1}\Big) \ket{\psi}\bra{\psi}\otimes\\
 \ket{\beta}\bra{\beta}
 \Big( \Ket{\mathcal{O}}\bra{\mathcal{O}}\otimes \Ket{0}\bra{0} + \Ket{\mathcal{O}}\bra{\mathcal{I}}\otimes \Ket{1}\bra{0}\Big)I_{2^{n}}\otimes \ket{0}\\
 +I_{2^{n}}\otimes \bra{1} \Big(\Ket{\mathcal{I}
 }\bra{\mathcal{I}
 }\otimes \Ket{I}\bra{I} + \Ket{\mathcal{O}}\bra{\mathcal{I}}\otimes \Ket{1}\bra{0}\Big) \ket{\psi}\bra{\psi}\otimes\\
 \ket{\beta}\bra{\beta}
 \Big( \Ket{\mathcal{I}}\bra{\mathcal{I}}\otimes \Ket{1}\bra{1} + \Ket{\mathcal{O}}\bra{\mathcal{I}}\otimes \Ket{1}\bra{0}\Big)I_{2^{n}}\otimes \ket{1}=
\end{multline*}
\begin{multline*}
\Ket{\mathcal{O}}\bra{\mathcal{O}}\ket{\psi}\braket{0|\beta}\braket{\beta|0}\bra{\psi}\Ket{\mathcal{O}}\bra{\mathcal{O}}+\Ket{\mathcal{O}}\bra{\mathcal{O}}\ket{\psi}\braket{0|\beta}\braket{\beta|1}\bra{\psi}\Ket{\mathcal{O}}\bra{\mathcal{I}}+\\
\Ket{\mathcal{I}}\bra{\mathcal{O}}\ket{\psi}\braket{1|\beta}\braket{\beta|0}\bra{\psi}\Ket{\mathcal{O}}\bra{\mathcal{O}}+
\Ket{\mathcal{I}}\bra{\mathcal{O}}\ket{\psi}\braket{1|\beta}\braket{\beta|1}\bra{\psi}\Ket{\mathcal{O}}\bra{\mathcal{I}}+\\
\Ket{\mathcal{I}}\bra{\mathcal{I}}\ket{\psi}\braket{1|\beta}\braket{\beta|1}\bra{\psi}\Ket{\mathcal{I}}\bra{\mathcal{I}}+\Ket{\mathcal{I}}\bra{\mathcal{I}}\ket{\psi}\braket{1|\beta}\braket{\beta|0}\bra{\psi}\Ket{\mathcal{I}}\bra{\mathcal{O}}+\\ \Ket{\mathcal{O}}\bra{\mathcal{I}}\ket{\psi}\braket{0|\beta}\braket{\beta|1}\bra{\psi}\Ket{\mathcal{I}}\bra{\mathcal{I}}+\Ket{\mathcal{O}}\bra{\mathcal{I}}\ket{\psi}\braket{0|\beta}\braket{\beta|1}\bra{\psi}\Ket{\mathcal{I}}\bra{\mathcal{O}}
\end{multline*}

\begin{multline*}
\braket{0|\beta}\braket{\beta|0}\Big(\Ket{\mathcal{O}}\bra{\mathcal{O}}\ket{\psi}\bra{\psi}\Ket{\mathcal{O}}\bra{\mathcal{O}}+
\Ket{\mathcal{O}}\bra{\mathcal{I}}\ket{\psi}\bra{\psi}\Ket{\mathcal{I}}\bra{\mathcal{O}}\Big)+\\
\braket{1|\beta}\braket{\beta|1}\Big(\Ket{\mathcal{I}}\bra{\mathcal{O}}\ket{\psi}\bra{\psi}\Ket{\mathcal{O}}\bra{\mathcal{I}}+
\Ket{\mathcal{I}}\bra{\mathcal{I}}\ket{\psi}\bra{\psi}\Ket{\mathcal{I}}\bra{\mathcal{I}}\Big)+\\
\braket{1|\beta}\braket{\beta|0}\Big(\Ket{\mathcal{I}}\bra{\mathcal{O}}\ket{\psi}\bra{\psi}\Ket{\mathcal{O}}\bra{\mathcal{O}}+
\Ket{\mathcal{I}}\bra{\mathcal{I}}\ket{\psi}\bra{\psi}\Ket{\mathcal{I}}\bra{\mathcal{O}}\Big)+\\
\braket{0|\beta}\braket{\beta|1}\Big(\Ket{\mathcal{O}}\bra{\mathcal{O}}\ket{\psi}\bra{\psi}\Ket{\mathcal{O}}\bra{\mathcal{I}}+
\Ket{\mathcal{O}}\bra{\mathcal{I}}\ket{\psi}\bra{\psi}\Ket{\mathcal{I}}\bra{\mathcal{I}}\Big)
\end{multline*}
\begin{multline*}
\braket{0|\beta}\braket{\beta|0}\Ket{\mathcal{O}}\Big(\bra{\mathcal{O}}\ket{\psi}\bra{\psi}\Ket{\mathcal{O}}+
\bra{\mathcal{I}}\ket{\psi}\bra{\psi}\Ket{\mathcal{I}}\Big)\bra{\mathcal{O}}+\\
\braket{1|\beta}\braket{\beta|1}\Ket{\mathcal{I}}\Big(\bra{\mathcal{O}}\ket{\psi}\bra{\psi}\Ket{\mathcal{O}}+
\bra{\mathcal{I}}\ket{\psi}\bra{\psi}\Ket{\mathcal{I}}\Big)\bra{\mathcal{I}}+\\
\braket{0|\beta}\braket{\beta|1}\Ket{\mathcal{O}}\Big(\bra{\mathcal{O}}\ket{\psi}\bra{\psi}\Ket{\mathcal{O}}+
\bra{\mathcal{I}}\ket{\psi}\bra{\psi}\Ket{\mathcal{I}}\Big)\bra{\mathcal{I}}+\\
\braket{1|\beta}\braket{\beta|0}\Ket{\mathcal{I}}\Big(\bra{\mathcal{O}}\ket{\psi}\bra{\psi}\Ket{\mathcal{O}}+
\bra{\mathcal{I}}\ket{\psi}\bra{\psi}\Ket{\mathcal{I}}\Big)\bra{\mathcal{I}}=
\end{multline*}

\begin{multline*}
\braket{0|\beta}\braket{\beta|0} I_{2^{n-1}}\otimes\Ket{0} tr_n\{\Ket{\psi}\bra{\psi}\}I_{2^{n-1}}\otimes\bra{0}+\\
\braket{1|\beta}\braket{\beta|1} I_{2^{n-1}}\otimes\Ket{1} tr_n\{\Ket{\psi}\bra{\psi}\}I_{2^{n-1}}\otimes\bra{1}+\\
\braket{0|\beta}\braket{\beta|1} I_{2^{n-1}}\otimes\Ket{0} tr_n\{\Ket{\psi}\bra{\psi}\}I_{2^{n-1}}\otimes\bra{1}+\\
\braket{1|\beta}\braket{\beta|0} I_{2^{n-1}}\otimes\Ket{1} tr_n\{\Ket{\psi}\bra{\psi}\}I_{2^{n-1}}\otimes\bra{0}=
\end{multline*}
\begin{multline*}
tr_n\{\Ket{\psi}\bra{\psi}\}\Big(\braket{0|\beta}\braket{\beta|0}\Ket{0}\bra{0}+\braket{1|\beta}\braket{\beta|1}\Ket{1}\bra{1}+\\
\braket{0|\beta}\braket{\beta|1}\Ket{0}\bra{1}+\braket{1|\beta}\braket{\beta|0}\Ket{1}\bra{0}\Big) =\\ tr_{n}\{\ket{\psi}\bra{\psi}\}\otimes\ket{\beta}\bra{\beta}
\;\;\;\;\;\square
\end{multline*}
\subsection{\label{partialtrace} Partial trace is sum of system when bath is measured in different states}
\begin{equation}
tr_{k}\{\Ket{\psi}\bra{\psi}\} = \sum_{i=0}^1 I_{2^{k-1}}\otimes\bra{i}\otimes I_{2^{N-k}}\Ket{\psi}\bra{\psi}I_{2^{k-1}}\otimes\Ket{i}\otimes I_{2^{N-k}}
\end{equation}
\begin{equation}
\nonumber
    I_{2^{k-1}}\otimes\Ket{i}\otimes I_{2^{N-k}} = P^i
\end{equation}
\begin{equation}
I_{2^{k-1}}\otimes\bra{i}\otimes I_{2^{N-k}} = (P^i)^\dagger
\end{equation}
\begin{equation}
    tr_{k}\{\Ket{\psi}\bra{\psi}\} = \sum_{i=0}^1 {P^i}^{\dagger}\Ket{\psi}\bra{\psi}P^i
\end{equation}

\(for\;\; b\in [0,2^{k-1}-1]\;\; and\;\; a\in [0,2^{N-k}-1]\)
\begin{equation}
    \hat P^i_{[l,m]} = \begin{cases}
                    1 \;\;if\;\; l = (2b+i)*2^{N-k}+a , m = b*2^{N-k}+a\\
                    0 \;\;else
               \end{cases}
\end{equation}
\begin{equation}
    \hat {P^i}^{\dagger}_{[l,m]} = \begin{cases}
                    1 \;\;if\;\; l = b*2^{N-k}+a , m = (2b+i)*2^{N-k}+a\\
                    0 \;\;else
               \end{cases}
\end{equation}

\(l = r*2^{N-k}+x \;\;\;\;\;\;j = y*2^{N-k}+z\;\;\; for\;\; y,r\in [0,k+1]\;\; and\;\; x,z\in [0,2^{N-k}]\)
\begin{equation}
    tr_{k}\{\Ket{\psi}\bra{\psi}\}_{l,j}= \sum_{i,d,e}\hat {P^i}^{\dagger}_{l,d}(\Ket{\psi}\bra{\psi})_{d,e}\hat P^i_{e,j} =
\end{equation}
\begin{equation}
\sum_{i=0}^1 \hat P^{i^{\dagger}}_{[r*2^{N-k}+x,(2r+i)*2^{N-k}+x]}(\Ket{\psi}\bra{\psi})_{[(2r+i)*2^{N-k}+x,(2y+i)*2^{N-k}+z]}\hat P^i_{[2(y+i)*2^{N-k}+z,y*2^{N-k}+z]}=
\end{equation}
\begin{equation}
(\sum_{i=0}^1 (\Ket{\psi}\bra{\psi})_{[(2r+i)*2^{N-k}+x,(2y+i)*2^{N-k}+z]})= \sum_{i=0}^1 N_i \Ket{\phi}^i_{r*2^{N-k}+x}N_i\bra{\phi}^i_{y*2^{N-k}+z} = 
\end{equation}
\begin{equation}
    \sum_{i=0}^1 N_i^2 \Ket{\phi}^i\bra{\phi}^i_{[r*2^{N-k}+x,y*2^{N-k}+z]} = \sum_{i=0}^1 N_i^2 \Ket{\phi}^l\bra{\phi}^i_{l,j} \;\;\;\;\;\square
\end{equation}
\newpage
\bibliography{name}
\end{document}